# Optical injection locking of white light cavity based superluminal lasers

Jacob Scheuer[1,*], Zifan Zhou[2], Yael Sternfeld[1,3], Michal Hazan[1], and Selim M. Shahriar[2,4]
[1]*School of Electrical & Computer Engineering, Tel-Aviv University, Tel-Aviv, Israel 6997801*
[2]*Department of Electrical & Computer Engineering, Northwestern University, Evanston, Illinois 60208, USA*
[3]*Department of Physics, University of California, Berkeley, California 94720. USA*
[4]*Department of Physics and Astronomy, Northwestern University, Evanston, Illinois 60208, USA*
[*]*kobys@tauex.tau.ac.il*

Optical injection locking is a simple method for synchronizing a laser to an external laser source by injecting part of the external laser light into the cavity of the other laser. This approach received significant interest due to its potential applications in telecommunications, precision metrology, and more. We study and analyze the steady-state and dynamic properties of optical injection locking in white light cavity (WLC) lasers. We derive the steady-state injection locking range of the WLC laser and the dynamic ("Adler") equation for the phase-difference between the injected signal and the locked laser output. The analysis shows that WLC lasers can exhibit an order of magnitude broader locking range and three order of magnitudes faster dynamics compared to conventional lasers with similar thresholds and free spectral ranges.

## I. INTRODUCTION

Optical injection locking (OIL) is a technique for synchronizing (in frequency and phase) a free-running laser ("target") to an external laser source ("master") by means of injecting part of the master laser light into the cavity of the target (see Fig. 1) [1]. When the lasing frequency of the master laser is sufficiently close to that of the free-running target laser, the latter will be forced to synchronize its frequency (and phase) with that of the master. In other words, the target would lase at the same frequency as the master with a fixed phase difference.

Fig. 1 shows a schematic of an integrated ring laser coupled to I/O waveguide by means of a directional coupler with power coupling coefficient $\kappa$. The small-signal roundtrip gain in the laser is assumed to be $g$. The amplitude of the injected signal is $E_1$ (with frequency $\omega$) and the corresponding output amplitude is $E_{out}$. The output power of the free running laser (for $E_1$=0) is assumed to be $E_0$ (with frequency $\omega_0$). Note that this scheme is slightly different than that presented in [1], leading to slightly different (though conceptually similar) results. A comprehensive analysis of injection locking of the scheme of Fig. 1 as well as dynamic effects and the derivation of the corresponding Adler equation [2] is given in the Appendix.

Locking a laser to an external source reference is an important building block for various applications [3]. Synchronized lasers can be used as local oscillators for coherent optical communications [4]-[6]. It can also be used for selecting and recovering optical carriers for coherent optical communications both in transmitter and receiver sides [7]-[11]. In addition to frequency and phase synchronization, OIL can be sued for enhancing the modulation bandwidth of directly modulated target lasers while reducing modulation chirp and relative intensity noise [12]-[16]. An important feature of OIL is the ability to reduce the linewidth of low-cost, powerful, lasers by injection locking them to a highly stable low-power laser. This is crucial for applications including coherent optical communication systems and microwave photonics. While linewidth reduction can be obtained by using an external cavity or by locking the laser to a stable reference with advanced electronics, these approaches are complex and expensive [17], [18]. On the other hand, OIL offers a relatively low-cost solution to this problem. On the downside, the mechanism that locks the lasing frequency of a laser to a small, injected signal may prove to be detrimental for some applications. A particular example of such a scenario is the ring laser gyroscope (RLG) [1], [19]. In an RLG, two counterpropagating lasers operate within the same ring cavity. When the cavity is rotated, the lasing frequencies of the two counterpropagating beams split, where the splitting is proportional to the rotation rate. This frequency splitting can be used for extracting the rotation rate (e.g. by beating the two beams). However, even a small amount backscattering from one beam to the other leads to mutual frequency lock-in and limiting the ability to detect small rotation rates. This "lock-in" phenomenon stems from the OIL mechanism which occurs when the difference between the two frequencies becomes small enough to place them within the locking range of each other.

A white light cavity (WLC) is a unique type of resonator which is designed to resonate over a broad, continuous range of frequencies [20], [21]. This is accomplished by introducing an intra-cavity element which provide a phase shift with a slope that is opposite in sign and equal in magnitude to that accumulated by light propagation through the "conventional" part of the structure [21]-[23]. Such a phase compensator can be realized by means of e.g., absorptive vapor cells [24] and lossy resonators [25]. When optical gain medium is introduced into a WLC the resulting system forms a unique laser scheme often denoted as a superluminal laser [26]-[31]. In a scenario where the gain is very broad and introduces negligible dispersion, we refer to such a system as a WLC laser. Such lasers have acquired increasing interest in the past decade due to their potential for realizing highly sensitive optical sensors and gyroscopes [26], [27], [31]-[33].

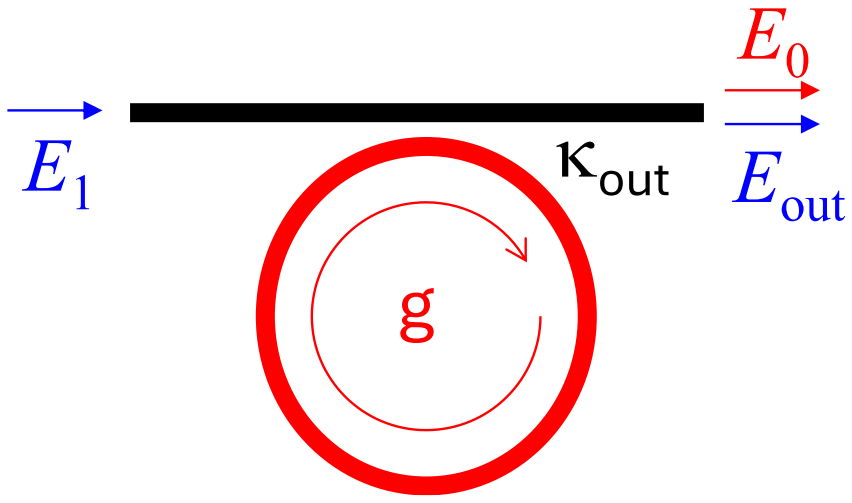

**Fig. 1:** Schematic of injection locking scheme of a ring laser

In this paper we investigate the injection locking phenomenon in a WLC laser. Because of its unique spectral properties, which differ substantially from those of conventional lasers, it can be expected that the WLC lasers exhibit different OIL properties and dynamics. Fig. 2 depicts a schematic of the WLC laser model analyzed in this paper. The scheme consists of two coupled ring resonators with coupling power coefficient $\kappa$. The right ring incorporates optical gain with roundtrip gain $g$ while the other (left) is assumed to be transparent. The transparent resonator is coupled to a waveguide by means of a directional coupler with power coupling coefficient $\kappa_{out}$. For simplicity, it is assumed that the two resonators are identical, i.e. they exhibit identical resonance frequencies. For free-running conditions, the amplitude of the emitted field from the waveguide is assumed to be $E_0$. Under external injection conditions, the amplitude of the injected field is $E_1$ and the corresponding output field is $E_{out}$. The amplitudes $a$ and $b$ correspond to the fields in the transparent resonator just above and below the output coupler while the amplitudes $c$ and $d$ indicate the fields in the active resonator below and above the coupler (see Fig. 2).

The rest of the paper is organized as follows: in Section II we analyze the WLC laser under external CW injection and obtain the frequency locking range. In Section III we derive the locking dynamic equation governing the WLC laser (i.e. the corresponding Adler equation) and obtain the locked laser phase shift. In Section IV we summarize and conclude.

## II. OIL OF A WLC LASER

In this section we analyze OIL in a WLC laser (see Fig. 2) in a similar manner to that in a conventional laser (Fig. 1) as outlined in the Appendix. We assume that without the injected signal (i.e. $E_1$=0) the laser is pumped above the threshold level and free-running with a frequency $\omega_0$ and output field amplitude of $E_0$. As for OIL in the conventional laser case, we consider the system for the injected signal as a regenerative amplifier [1] and that the gain is clamped to its threshold level $g = g_{th}$. As noted above, we also assume that the gain profile is relatively broad (spectrally) such that we can assume the roundtrip gain in the vicinity of $\omega_0$ to be frequency independent. Finally, as we focus on the fundamental and conceptual properties of OIL in a WLC laser, we assume that the linewidth enhancement factor (which is common in semiconductor lasers and also known as the Henry factor [34]) is negligible. In this context, we note that very low linewidth enhancement factors are activable in properly designed InAs/GaAs quantum dot lasers [35].

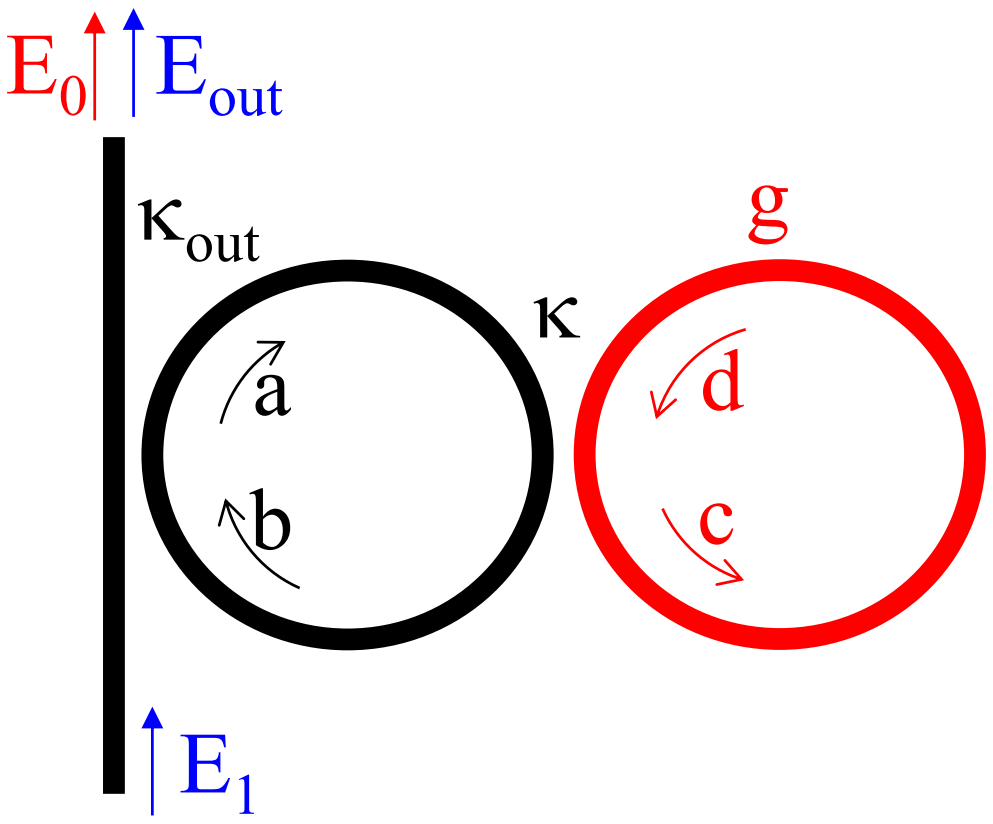

**Fig. 2:** Schematic of a WLC laser model. The right (red) ring resonator corresponds to the main cavity incorporating optical gain. The left (black) ring resonator and the waveguide together realize the phase compensation component.

In steady-state, the relations between the various amplitudes of the field in the device satisfy the following relations:

$$\begin{aligned} d &= ce^{-i\phi}g \\ be^{+i\frac{\phi}{2}} &= ae^{-i\frac{\phi}{2}}\sqrt{1-\kappa} + i\sqrt{\kappa}d \\ c &= i\sqrt{\kappa}ae^{-i\frac{\phi}{2}} + \sqrt{1-\kappa}d \\ &= i\sqrt{\kappa}ae^{-i\frac{\phi}{2}} + c\sqrt{1-\kappa}e^{-i\phi}g \\ a &= b\sqrt{1-\kappa_{out}} + i\sqrt{\kappa_{out}}E_1 \\ E_{out} &= E_1\sqrt{1-\kappa_{out}} + i\sqrt{\kappa_{out}}b \end{aligned} \tag{1}$$

Note that $\phi$ is the round-trip phase at the frequency of the *injected signal* at frequency $\omega$. Solving Eq. (1) yields the following expression for the output field:

$$\frac{E_{out}}{E_1} = \frac{\sqrt{1-\kappa_{out}} - \sqrt{(1-\kappa_{out})(1-\kappa)}ge^{-i\phi} - \sqrt{1-\kappa}e^{-i\phi} + ge^{-2i\phi}}{1-\sqrt{1-\kappa}e^{-i\phi}\left(g+\sqrt{1-\kappa_{out}}\right) + g\sqrt{1-\kappa_{out}}e^{-2i\phi}} \tag{2}$$

To continue, we need to obtain the lasing threshold gain for the free-running laser case and to express the average group index in terms of $\kappa$ and $\kappa_{out}$. The threshold condition is obtained by setting the denominator of Eq. (2) to zero (i.e. a pole in the transmission function). This equation includes both real and imaginary parts that should be set to zero:

$$\begin{aligned} 1-\sqrt{1-\kappa}\left(g_{th}+\sqrt{1-\kappa_{out}}\right)\cos\phi \\ +g_{th}\sqrt{1-\kappa_{out}}\cos 2\phi=0 \\ \sqrt{1-\kappa}\left(g_{th}+\sqrt{1-\kappa_{out}}\right)\sin\phi \\ -g_{th}\sqrt{1-\kappa_{out}}\sin 2\phi=0 \end{aligned} \quad (3)$$

Eq. (3) exhibits two solutions: $\sin\phi=0$ and $\sqrt{1-\kappa}\left(g_{th}+\sqrt{1-\kappa_{out}}\right)=2g_{th}\sqrt{1-\kappa_{out}}\cos\phi$. The first solution corresponds to the free-running laser lasing at $\omega_0$ (the resonance frequency of each ring). The second solution (which consists of two possible frequencies $\pm\Delta\omega$) depends on the specific parameters of the structure. Particularly, this solution might not exist if $|\cos\phi|$ turns out to be larger than 1. This solution exists if the group index induced by the phase component (the passive resonator coupled to the waveguide) over-compensates the roundtrip phase in the active cavity (leading to a negative average group index). This scenario might occur if $\kappa_{out}$ is too small (see e.g.[36]). As this is an undesired scenario, we assume that the single solution ($\omega=\omega_0$) holds. Substituting $\phi=0$ to Eq. (3) leads to the threshold gain:

$$g_{th}=\frac{1-\sqrt{1-\kappa}\sqrt{1-\kappa_{out}}}{\sqrt{1-\kappa}-\sqrt{1-\kappa_{out}}} \quad (4)$$

Clearly, the threshold gain depends on the choice of the coupling parameters $\kappa$ and $\kappa_{out}$. However, for sensing applications it is advantageous to operate at a point where the group delay in the laser vanishes (the WLC condition). In the scheme of Fig. 2, this requirement imposes the following relation between $\kappa$ and $\kappa_{out}$ [27]:

$$\sqrt{1-\kappa}=\frac{2}{\sqrt{1-\kappa_{out}}+1/\sqrt{1-\kappa_{out}}} \quad (5)$$

Introducing (5) into (4) yields the threshold gain:

$$g_{th}\sqrt{1-\kappa_{out}}=1 \quad (6)$$

Note that Eq. (6) is identical to Eq. (22) in the Appendix, i.e., in the WLC laser case the threshold condition is identical to that of the simple ring laser. Introducing Eqs. (5) and (6) into Eq. (2) yields the following transmission function of the WLC laser to an external signal $E_1$:

$$\begin{aligned} &\frac{E_{out}}{E_1} \\ &=\frac{(1-\kappa_{out})(2-\kappa_{out})-4(1-\kappa_{out})e^{-i\phi}-(2-\kappa_{out})e^{-2i\phi}}{(2-\kappa_{out})\sqrt{1-\kappa_{out}}(1-e^{-i\phi})^2} \end{aligned} \quad (7)$$

As for the conventional ring laser OIL, we obtain the locking rage by setting $E_{out}=E_0$ and approximating $e^{-i\phi}\approx 1-i\phi$. Substituting these into Eq. (7) yields:

$$\frac{E_0}{E_1}=-\frac{\kappa_{out}}{\sqrt{1-\kappa_{out}}}\cdot\frac{\kappa_{out}-2i\phi_{lock}}{(2-\kappa_{out})\phi_{lock}^2} \quad (8)$$

Taking the absolute value of Eq. (8) and solving for $\phi_{lock}$ yields:

$$\begin{aligned} \phi_{lock}^2 &=\frac{\kappa_{out}^2}{(2-\kappa_{out})\sqrt{1-\kappa_{out}}\cdot\left|\frac{E_0}{E_1}\right|-2} \\ &\approx\frac{\kappa_{out}^2}{(2-\kappa_{out})\sqrt{1-\kappa_{out}}\cdot\left|\frac{E_0}{E_1}\right|} \end{aligned} \quad (9)$$

The approximation in Eq. (9) stems from the assumptions that $\kappa_{out}$ is relatively small (e.g. $\kappa_{out}<\frac{1}{2}$) and that $\left|\frac{E_1}{E_0}\right|\ll 1$ (small injection). From Eq. (9), the OIL range can be readily extracted (similar to the simple ring laser case):

$$\Delta\omega_{lock}^2\approx\frac{\Delta\nu_{FSR}^2\cdot\kappa_{out}^2}{(2-\kappa_{out})\sqrt{1-\kappa_{out}}}\sqrt{\frac{I_1}{I_0}} \quad (10)$$

where $I_0=|E_0|^2$ and $I_1=|E_1|^2$. Note that an important result of Eq. (10) is that the locking range is proportional to the square root of the amplitudes ratio and not to the ratio itself, just as in the conventional ring laser case (Eq. (26) in the Appendix).

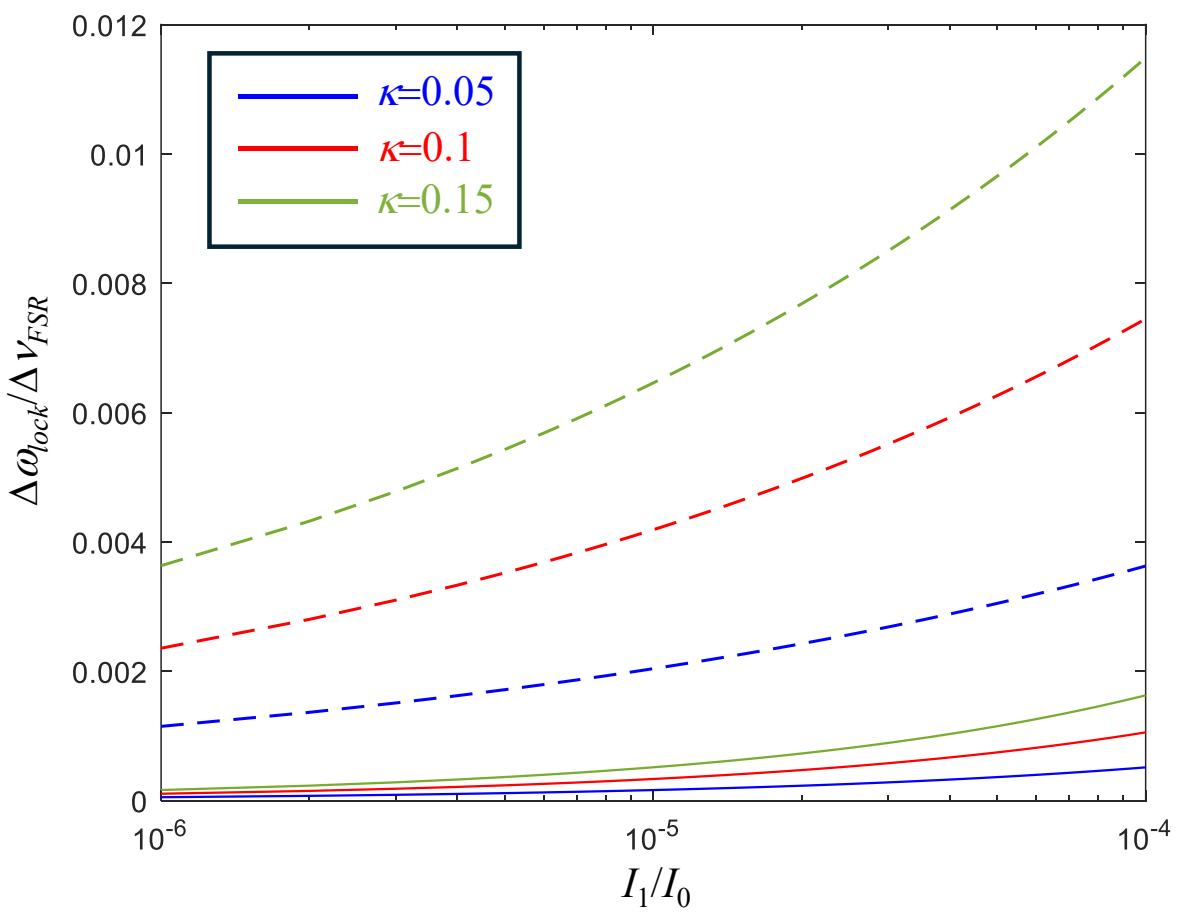

**Fig. 3:** locking range vs. intensities ratio for various output coupling coefficients in a regular laser (solid) and WLC laser (dashed). Colors correspond to different output coupling values.

Fig. 3 compares the locking range as a function of $I_1/I_0$ of the regular ring laser (solid lines) and the WLC laser (dashed). To keep a general description, the locking range is normalized to the Free Spectral Range (FSR) of the resonators. The colors correspond to different output coupling levels. There are several phenomena that can be observed in Fig. 3. First, the locking range of the WLC laser (under the same output coupling conditions which correspond to identical lasing threshold) is significantly larger than that of the conventional laser. Second, as the power of the injected signal is increased, the locking range increases as well (as expected) but the ratio of the locking ranges decreases. Finally, increasing the output coupling coefficient yields a larger locking range. Using Eq. (10) and Eq. (26) in the Appendix, we find that the ratio between the locking ranges of the WLC laser and the conventional ring laser is given by:

$$\frac{\Delta\omega_{lock}^{WLC}}{\Delta\omega_{lock}^{Conv}} = \frac{\sqrt[4]{1-\kappa_{out}}}{\sqrt{2-\kappa_{out}}}\sqrt[4]{I_0/I_1} \tag{11}$$

From Eq. (11), it is clear that the expansion of the locking range in the WLC laser is inversely proportional to the square root of the ratio between the field amplitude of the injected signal and that of the free-running laser. This is clearly seen in Fig. 3. Interestingly, the coefficient multiplying the powers ratio in Eq. (11) is weakly dependent on the output coupling coefficient and remains very close to $\sim 1/\sqrt{2}$ for $\kappa_{out} < 0.4$. As the majority of practical laser devices obey this condition, we can approximate Eq. (11) by:

$$\frac{\Delta\omega_{lock}^{WLC}}{\Delta\omega_{lock}^{Conv}} \approx \frac{1}{\sqrt{2}\cdot\sqrt[4]{I_1/I_0}} \tag{12}$$

Thus, a direct impact of the WLC laser configuration is that it allows for a substantially broader locking range (compared to conventional lasers with same threshold level), especially for small, injected signals.

## III. OIL DYNAMICS IN WLC LASERS

In this section we obtain the dynamic equation for the injection locking of the WLC laser (i.e. the corresponding "Adler" equation), following an approach which is similar to that presented in the Appendix. We assume that the frequency of the injected signal is close to the resonance of the free-running laser so that we can approximate $e^{-i\phi} \approx 1 - i\phi$ where $\phi = (\omega - \omega_0)/\Delta\nu_{FSR}$ (the roundtrip phase in the resonator). Substituting this approximation into Eq. (7) yields:

$$\frac{\sqrt{1-\kappa_{out}}(2-\kappa_{out})}{\kappa_{out}^2\Delta\nu_{FSR}^2}(\omega-\omega_0)^2E_{out} = \left[\frac{2i(\omega-\omega_0)}{\Delta\nu_{FSR}\kappa_{out}} - 1\right]E_1 \tag{13}$$

Let us define $\alpha^2 = \frac{\sqrt{1-\kappa_{out}}(2-\kappa_{out})}{\kappa_{out}}$, and perform an inverse Fourier transform of (13):

$$-\frac{d^2E_{out}}{dt^2} - 2i\omega_0\frac{dE_{out}}{dt} + \omega_0^2E_{out} = -\frac{2\Delta\nu_{FSR}}{\alpha^2}\frac{dE_1}{dt} - \frac{2i\omega_0\Delta\nu_{FSR}}{\alpha^2}E_1 - \frac{\kappa_{out}\Delta\nu_{FSR}^2}{\alpha^2}E_1 \tag{14}$$

Next, we write $E_{out}$ and $E_1$ as slowly varying amplitudes around frequency $\omega_1$ (the frequency of the injected signal): $E_{out} = A_{out}(t)e^{-i\omega t}, E_1 = A_1(t)e^{-i\omega t}$. Introducing this into Eq. (14) yields an equation of the slowly varying envelope $A_{out}$:

$$\frac{d^2A_{out}}{dt^2} - 2i(\omega_1-\omega_0)\frac{dA_{out}}{dt} - (\omega_1-\omega_0)^2A_{out} = \frac{\kappa_{out}\Delta\nu_{FSR}^2}{\alpha^2}A_1 - \frac{2i(\omega_1-\omega_0)\Delta\nu_{FSR}}{\alpha^2}A_1 + \frac{2\Delta\nu_{FSR}}{\alpha^2}\frac{dA_1}{dt} \tag{15}$$

The last two terms in the right-hand side (RHS) of Eq. (15) can be neglected because of the following reasons: $A_1$ is assumed to be almost constant and, hence, its derivative is small. It is also assumed that $|\omega_0 - \omega_1| \ll \Delta\nu_{FSR}$ and therefore the first term in the RHS of Eq. (15) is substantially larger than the second term. Next, we separate the magnitude and phase of $A_{out}$ and $A_1$: $A_{out} = \bar{A}_{out}(t)e^{-i\theta(t)}$; $A_1 = \bar{A}_1(t)e^{-i\theta_1(t)}$. Introducing these definitions into (15) yields:

$$\begin{aligned}\frac{d^2\bar{A}_{out}}{dt^2} - 2i\frac{d\bar{A}_{out}}{dt}\frac{d\theta}{dt} - i\bar{A}_{out}\frac{d^2\theta}{dt^2} - \bar{A}_{out}\left(\frac{d\theta}{dt}\right)^2 - 2i(\omega-\omega_0)\left[\frac{d\bar{A}_{out}}{dt} - i\bar{A}_{out}\frac{d\theta}{dt}\right] - (\omega-\omega_0)^2\bar{A}_{out} \\ = \frac{\kappa_{out}\Delta\nu_{FSR}^2}{\alpha^2}\bar{A}_1e^{-i(\theta_1-\theta)} = \frac{\kappa_{out}\Delta\nu_{FSR}^2}{\alpha^2}\bar{A}_1\cdot[\cos(\theta_1-\theta) - i\sin(\theta_1-\theta)]\end{aligned} \tag{16}$$

We separate Eq. (16) into its real and imaginary parts and assume that within the locking range the magnitude of the output field does not vary significantly, i.e. $\bar{A}_{out} \approx A_0$ [1]. The real part of Eq. (16) yields:

$$\left(\frac{d\theta}{dt}+(\omega-\omega_0)\right)^2=-\Delta\omega_{lock}^2\cos(\theta_1-\theta) \qquad (17)$$

where $\Delta\omega_{lock}^2$ is defined in Eq. (10). Eq. (17) yields a first order differential equation for the output phase:

$$\frac{d\theta}{dt}=-(\omega-\omega_0)\pm\Delta\omega_{lock}\sqrt{-\cos(\theta_1-\theta)} \qquad (18)$$

Note that there is a steady state solution for $\theta$ as long as $|\omega-\omega_0|\leq\Delta\omega_{lock}$. The choice of sign in Eq. (18) stems for the detuning of $\omega$ from $\omega_0$ (plus for positive detuning, minus for negative detuning). As for OIL of a conventional ring laser, in steady state there is a shift between the phase of the output (locked) field and that of the injected signal. This phase shift is given by:

$$\cos(\Delta\theta)|_{steady\ state}=-\frac{(\omega-\omega_0)^2}{\Delta\omega_{lock}^2} \qquad (19)$$

where $\Delta\theta=\theta_1-\theta$ is the phase shift between $A_1$ and $A_{out}$. Note that because of the cosine function in the RHS of Eq. (19) there are two possible solutions for the steady-state phase-difference, for a given detuning. The specific solution (positive and negative $\Delta\theta$) depends on the initial conditions $\Delta\theta(t=0)$.

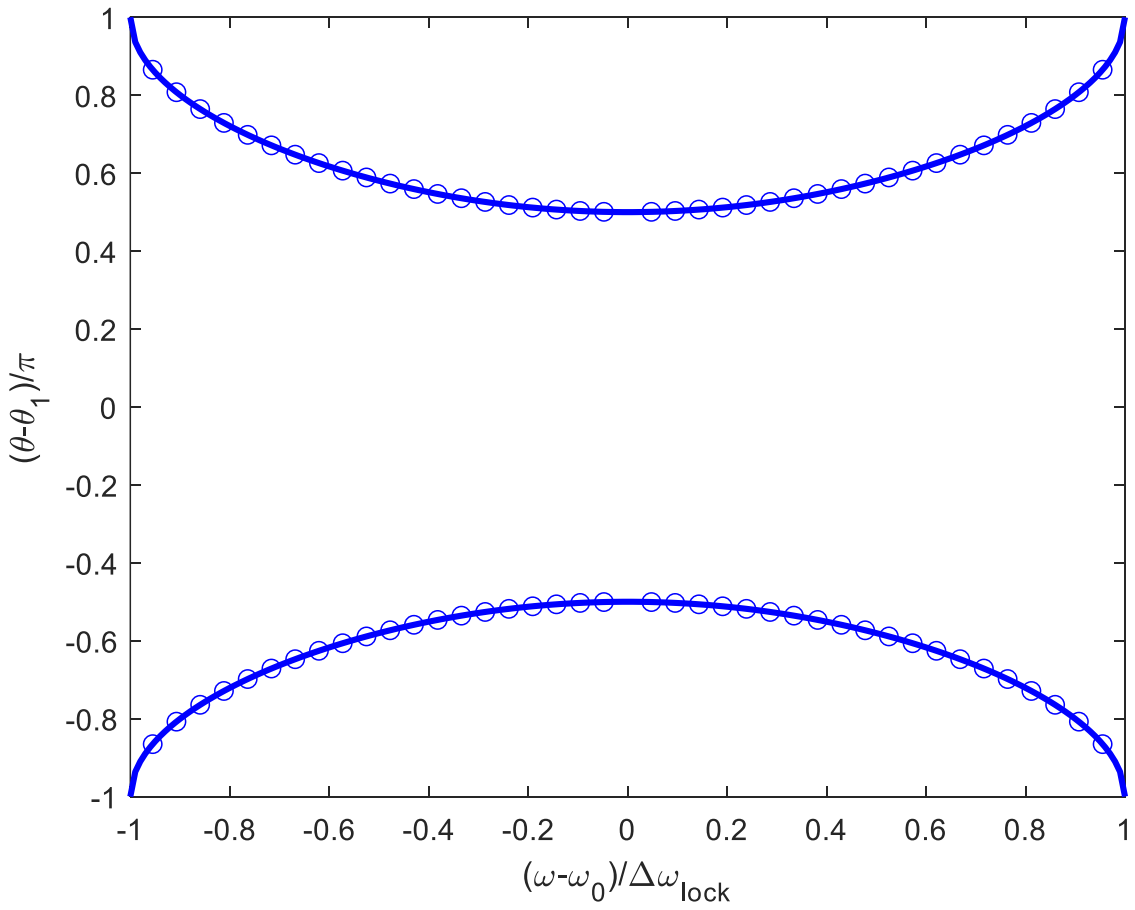

**Fig. 4**: Steady state output phase as a function of the frequency detuning. The solid lines correspond to the solutions of (19). circles indicate steady-state solutions obtained by numerically integrating (18).

Fig. 4 shows the steady-state phase difference between the output and that of the injected signal as a function of the detuning. The circles indicate steady-state solutions obtained by numerically integrating Eq. (18) (see below for details). Comparing Eq. (19) to the corresponding relation in a conventional laser (Eq. (33) in the Appendix) leads to several important observations. First, the steady state phase shift in the WLC laser case is $\frac{\pi}{2}<|\Delta\theta|<\pi$ in contrast to $-\frac{\pi}{2}<\Delta\theta<\frac{\pi}{2}$ in the conventional laser case. This is because of the minus sign under the square root in the RHS of (18) which requires $\cos(\Delta\theta)$ to be negative. Second, as mentioned above, there are two possible steady-state phase differences for a given detuning. Finally, note that near zero detuning (i.e., when the frequency of the injected signal is very close to that of the free-running laser), the steady-state phase difference is $\Delta\theta\to\pm\pi$ while for the conventional laser $\Delta\theta\to 0$. We believe that this stems from the fact that the signal is not injected directly into the laser cavity but rather through an add-drop filter.

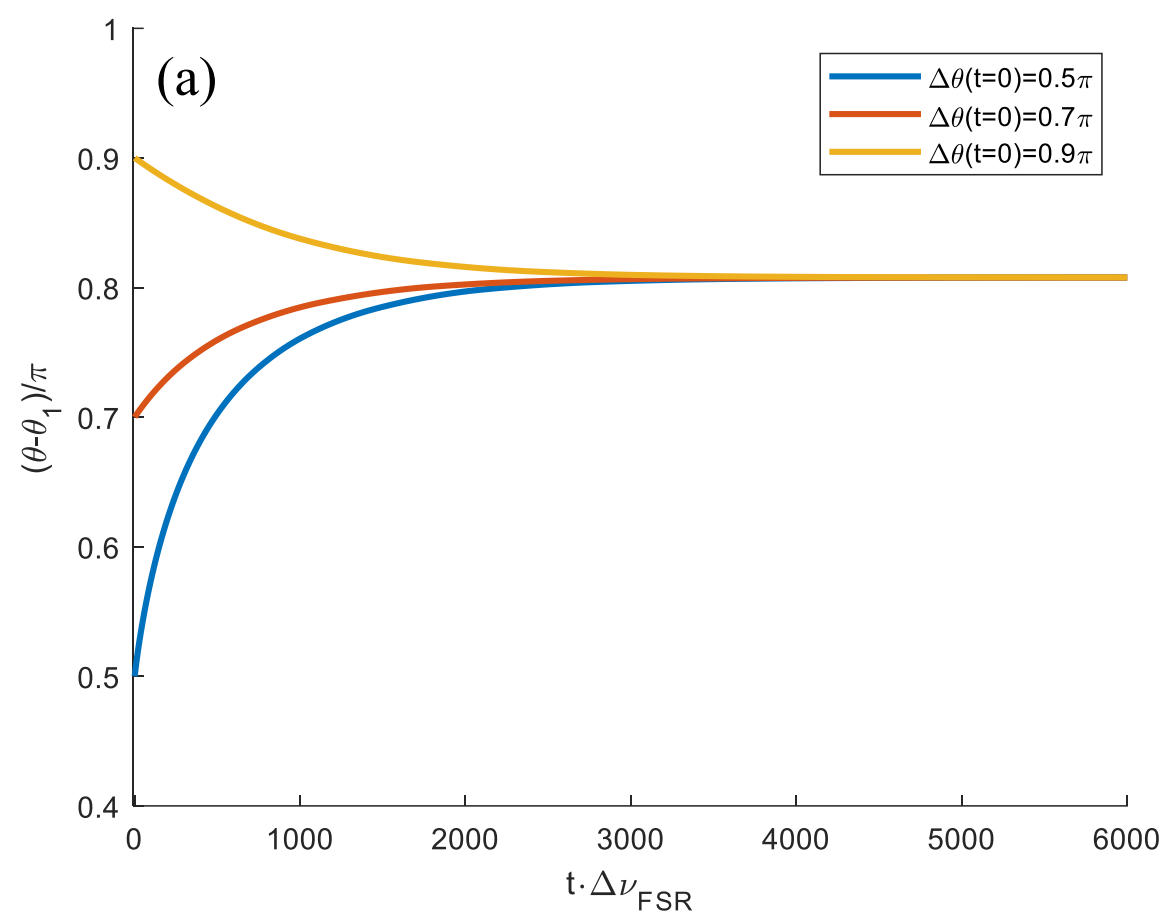

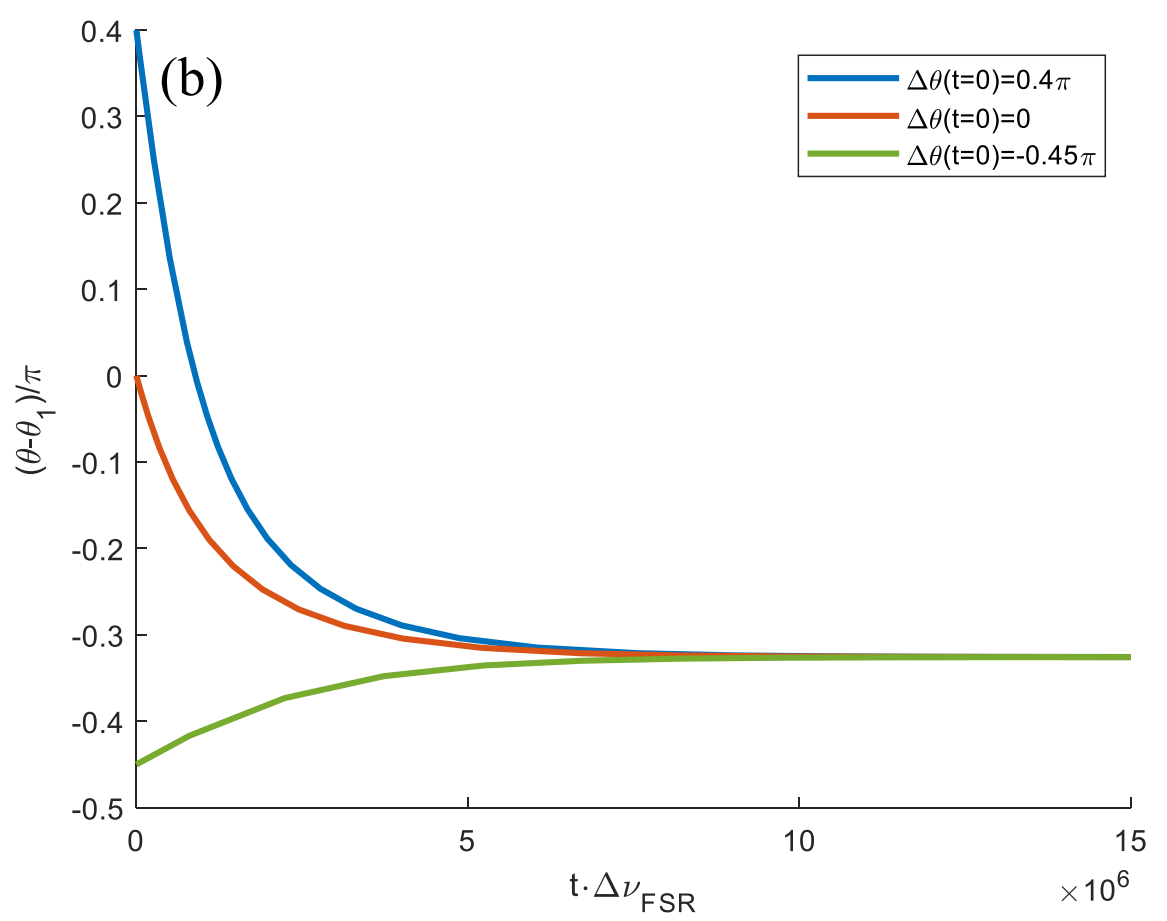

**Fig. 5**: Phase difference locking dynamics for (a) WLC laser and (b) conventional ring laser for various initial phase-difference values.

Fig. 5 compares the phase-difference locking dynamics in a WLC laser and in a conventional ring laser for various initial phase differences. The output coupling coefficient and frequency detuning are, respectively, $K_{out}=0.1$ and $\omega-\omega_0=0.9\cdot\Delta\omega_{lock}$ (note that the locking ranges of the lasers are different). The time axis in Fig. 5 is normalized to the roundtrip time in the individual rings ($T_{RT}=1/\Delta\nu_{FSR}$).

A striking result of Fig. 5 is that the locking time of the WLC laser is substantially shorter than that of the equivalent conventional laser (by approximately a factor of 2500). Thus, not only is the locking range of the WLC laser larger, but the locking dynamics is also substantially faster. Such properties can be highly useful for various applications, particularly telecommunications. We attribute the faster dynamics to the broader WLC resonance [41]. We again note that the broader resonance is not accompanied by larger cavity losses as the threshold levels of both lasers are identical, as manifested in Eqs. (6) and (22).

## IV. SUMMARY AND CONCLUSIONS

We have studied the locking range and dynamics of OIL in WLC lasers. We find that compared to an equivalent conventional laser (same threshold level and FSR) the WLC laser exhibits a broader locking range and faster dynamics. These properties render OIL in WLC attractive for applications requiring wider locking ranges and faster response. Alternatively, it allows for obtaining locking under lower injection level. On the other hand, for some applications, such as the RLG, the broader locking range might increase the minimal detectable rotation rate due to the frequency lock-in phenomenon. Because of the equivalence between WLCs and exceptional points (EPs) in optical systems [36], this may also be the case for *PT*-Symmetry based optical gyros (see also below) [37]-[40].

The structure studied here utilizes an add-drop filter to obtain a phase response that compensates for the propagation phase accumulation in the right cavity of Fig. 2. Under these conditions, the group index at $\omega=\omega_0$ is 0. As noted above, such a structure is equivalent to a *PT*-symmetric system with a second order EP. Utilizing the equivalence between the WLC condition and EPs in *PT*-Symmetry it is possible to design lasers operating at higher order EPs by introducing additional resonators [41]. Such lasers are expected to exhibit an even broader OIL range and faster locking dynamics.

Finally, as noted in the introduction, the presented model includes several assumptions, which might introduce some limitations. First, it is assumed that the gain is broad compared to the cavity linewidth. Second, the phase component needed for the WLC structure consists of an "artificial" absorber which is implemented by a cavity. This is unlike other implementations that utilize, e.g., Raman absorption and/or gain combinations [24], [26], [29]-[33]. The main impact of using an "artificial" absorber, in contrast to the Raman gain/loss approaches, is that the phase response of that absorber is independent of the laser intensity. Third, the device studied here operates under the WLC condition where the group index approaches zero. While this operation point simplifies the analysis, it is only a single point in the operation range of superluminal lasers, which is often difficult to achieve. Lasers are characterized as superluminal as long as $n_g<1$, which constitute a wide operation range that may exhibit different steady-state and dynamical properties than those described here. Fourth, it is assumed that the gain dynamics is substantially faster than that of the optical fields and that it can be adiabatically eliminated. This assumption can be justified for sufficiently high-Q cavities (small output coupling). Other implementations of WLC lasers may exhibit different steady-state and dynamic properties than those discussed here. However, the study of such implementation is beyond the scope of this paper and will be presented in a future publication.

### ACKNOWLEDGMENT

The authors are grateful to David D. Smith for useful discussions and insightful comments. This work has been supported by AFOSR (FA9550-18-01–0401, FA9550-21-C-0003 and FA9550-23-1–0617), NASA (80NSSC22CA052), Defense Security Cooperation Agency (PO4441028735), Israeli MOD (4441185451), and Israeli Innovation Authority (4692/1).

The authors declare no conflicts of interest.

## APPENDIX

To complement the analysis in this paper, the appendix derivates the locking range and the phase dynamics equation (i.e. the Adler equation) for the conventional ring laser shown schematically in Fig. 1. Referring to this figure, the input field of the injected signal is $E_1 = \sqrt{I_1}$, the output field at the frequency of the injected signal is $E_{out} = \sqrt{I_{out}}$, and the field of the free-running laser is $E_0 = \sqrt{I_0}$. $I_0$, $I_1$ and $I_{out}$ are the corresponding intensities of the free-running laser, the injected signal and the output (locked) signal, respectively. The relation between $I_{out}$ and $I_1$ is simply given by:

$$\frac{E_{out}}{E_1} = \frac{\sqrt{1-\kappa_{out}} - ge^{-i\phi}}{1-\sqrt{1-\kappa_{out}}ge^{-i\phi}} \tag{20}$$

$\phi = 2\pi R n_{eff}(\omega)\cdot\omega/c$ and $g$ are, respectively the roundtrip phase and gain in the cavity. If we are interested in the vicinity of a specific resonance frequency $\omega_0$ of the cavity, then the roundtrip phase can also be written as $\phi = 2\pi R n_{eff}(\omega)\cdot(\omega-\omega_0)/c = 2\pi R n_{eff}(\omega)\cdot\Delta\omega/c = \Delta\omega/\Delta\nu_{FSR}$. $\Delta\nu_{FSR}$ is the FSR of the cavity. The locking range can be found by setting $E_{out} = \sqrt{I_0}$ in Eq. (20), leading to:

$$\frac{I_{out}}{I_1} = \left|\frac{E_{out}}{E_1}\right|^2 = \frac{1-\kappa_{out}+g^2-2g\sqrt{1-\kappa_{out}}\cos\phi}{1+(1-\kappa_{out})g^2-2g\sqrt{1-\kappa_{out}}\cos\phi} \tag{21}$$

Under lasing conditions, the gain is clamped to its threshold level which is given by:

$$g_{th}\sqrt{1-\kappa_{out}} = 1 \tag{22}$$

Eq. (22) stems from the fact that the roundtrip loss in the system presented in Fig. 1 is caused by the output coupler. Introducing (22) into (21) yields:

$$\frac{I_{out}}{I_1} = \frac{1-\kappa_{out}+\frac{1}{1-\kappa_{out}}-2\cos\phi}{2(1-\cos\phi)} \tag{23}$$

Substituting $1-\cos\phi = 2\sin^2\phi/2$ and introducing the locking condition, $I_{out}$=$I_0$, yields:

$$4\sin^2\phi_{lock}/2 = \frac{\kappa_{out}^2}{1-\kappa_{out}}\frac{1}{\frac{I_0}{I_1}-1} \tag{24}$$

Assuming that the locking range is not large (i.e. $\sin\phi \approx \phi$) and recalling that $I_0 \gg I_1$ yields:

$$\phi_{lock}^2 \approx \frac{\kappa_{out}^2}{1-\kappa_{out}}\frac{I_1}{I_0} \tag{25}$$

Therefore, the locking range is given by $-\Delta\omega_{lock} < \Delta\omega < \Delta\omega_{lock}$ where:

$$\Delta\omega_{lock} \approx \frac{\kappa_{out}\cdot\Delta\nu_{FSR}}{\sqrt{1-\kappa_{out}}}\sqrt{\frac{I_1}{I_0}} \tag{26}$$

In the good cavity limit This expression is equivalent to the locking range reported in [1], $\Delta\omega_{lock} \approx \frac{\omega_0}{Q_c}\sqrt{I_1/I_0}$ .

Next, we consider the dynamics of the output field (the Adler equation) in this case. Introducing the threshold gain condition into Eq. (20) yields:

$$\frac{E_{out}}{E_1} = \frac{\sqrt{1-\kappa_{out}} - \frac{1}{\sqrt{1-\kappa_{out}}}e^{-i\phi}}{1-\ e^{-i\phi}} \tag{27}$$

Introducing the approximation $e^{-i\phi} \approx 1-i\phi$, where $\phi = (\omega-\omega_0)/\Delta\nu_{FSR}$, and rearranging Eq. (27) leads to:

$$\begin{aligned} i\sqrt{1-\kappa_{out}}(\omega-\omega_0)E_{out} &= -\kappa_{out}\Delta\nu_{FSR}E_1 \\ &+ i(\omega-\omega_0)E_1 \end{aligned} \tag{28}$$

The inverse Fourier transform of Eq. (8) where $-i\omega \Leftrightarrow \frac{d}{dt}$ yields:

$$\begin{aligned} i\sqrt{1-\kappa_{out}}\omega_0 E_{out} + \sqrt{1-\kappa_{out}}\frac{dE_{out}}{dt} \\ = \kappa_{out}\Delta\nu_{FSR}E_1 + i\omega_0 E_1 + \frac{dE_1}{dt} \end{aligned} \tag{29}$$

The next step is to approximate $E_{out}$ and $E_1$ as slowly varying amplitude envelopes around frequency $\omega$. The reason is that we are interested in the locking dynamics where in steady state both $E_{out}$ and (obviously) $E_1$ are oscillating at frequency $\omega$: $E_{out} = A_{out}(t)e^{-i\omega t}, E_1 = A_1(t)e^{-i\omega t}$. Substituting into Eq. (29) yields:

$$\begin{aligned} i\sqrt{1-\kappa_{out}}A_{out}(\omega_0-\omega) + \sqrt{1-\kappa_{out}}\frac{dA_{out}}{dt} \\ = \kappa_{out}\Delta\nu_{FSR}A_1 + i(\omega_0-\omega)A_1 \\ + \frac{dA_1}{dt} \end{aligned} \tag{30}$$

The last two terms can be neglected because the amplitude of the injected signal is assumed to be almost fixed, and because it can be assumed that the locking range is much smaller than the FSR, $|\omega_0-\omega_1| \ll \Delta\nu_{FSR}$. Finally, we write the (complex) amplitudes as $A_{out} = \bar{A}_{out}(t)e^{-i\theta(t)}$; $A_1 = \bar{A}_1(t)e^{-i\theta_1(t)}$. Substituting these definitions into (30) yields:

$$\frac{d\bar{A}_{out}}{dt} - i\frac{d\theta}{dt}\bar{A}_{out} = i(\omega-\omega_0)\bar{A}_{out} + \frac{\kappa_{out}\Delta\nu_{FSR}}{\sqrt{1-\kappa_{out}}}\bar{A}_1 e^{-i(\theta_1-\theta)} \tag{31}$$

Separating Eq. (31) into real and imaginary parts yields master equations for the amplitude $\bar{A}_{out}$ and the phase $\theta$. The dynamic of the output intensity is often considered less important as it is determined mainly by the saturation level of the gain in the cavity. Therefore, it is common to approximate $\bar{A}_{out} \approx \sqrt{I_0}$ [1]. Furthermore, recall that $\bar{A}_1 = \sqrt{I_1}$. From the imaginary part of Eq. (31) we obtain the evolution equation for the phase of the output field:

$$\frac{d\theta}{dt} = -(\omega-\omega_0) + \frac{\kappa_{out}\Delta\nu_{FSR}}{\sqrt{1-\kappa_{out}}}\sqrt{\frac{I_1}{I_0}}\sin(\theta_1-\theta) = -(\omega-\omega_0) + \Delta\omega_{lock}\sin(\theta_1-\theta) \tag{32}$$

It should be noted that Eq. (32) agrees with the Adler equation as shown in [1]. In steady-state we can set $d\theta/dt = 0$ and obtain the phase difference between in the injected signal and the laser output:

$$\sin(\theta_1-\theta)|_{steady\ state} = \frac{(\omega-\omega_0)}{\Delta\omega_{lock}} \tag{33}$$